\documentclass[sigconf]{acmart}

\AtBeginDocument{%
  \providecommand\BibTeX{{%
    \normalfont B\kern-0.5em{\scshape i\kern-0.25em b}\kern-0.8em\TeX}}}

\setcopyright{acmcopyright}
\copyrightyear{2022}
\acmYear{2022}
\acmDOI{10.1145/3505284.3529959}

\acmConference[IMX '22]{ACM International Conference on Interactive Media Experiences}{June 22--24, 2022}{Aveiro, JB, Portugal}
\acmBooktitle{ACM International Conference on Interactive Media Experiences (IMX '22), June 22--24, 2022, Aveiro, JB, Portugal}
\acmPrice{15.00}
\acmISBN{978-1-4503-9212-9/22/06}

\begin{document}

\title{CalmResponses: Displaying Collective Audience Reactions in Remote Communication}

\author{Kiyosu Maeda}
\orcid{}
\authornotemark[1]
\affiliation{%
  \institution{University of Tokyo}
  \streetaddress{}
  \city{Tokyo}
  \state{}
  \country{Japan}
  \postcode{}
}
\email{kiyosu775@g.ecc.u-tokyo.ac.jp}

\author{Riku Arakawa}
\affiliation{%
  \institution{Carnegie Mellon University}
  \city{Pittsburgh}
  \country{USA}}
\email{rarakawa@cs.cmu.edu}

\author{Jun Rekimoto}
\affiliation{%
  \institution{University of Tokyo}
  \city{Tokyo}
  \country{Japan}
  \institution{Sony CSL Kyoto}
  \city{Kyoto}
  \country{Japan}}
\email{rekimoto@acm.org}

\renewcommand{\shortauthors}{Maeda et al.}

\begin{teaserfigure}
\centering
\includegraphics[width=15cm]{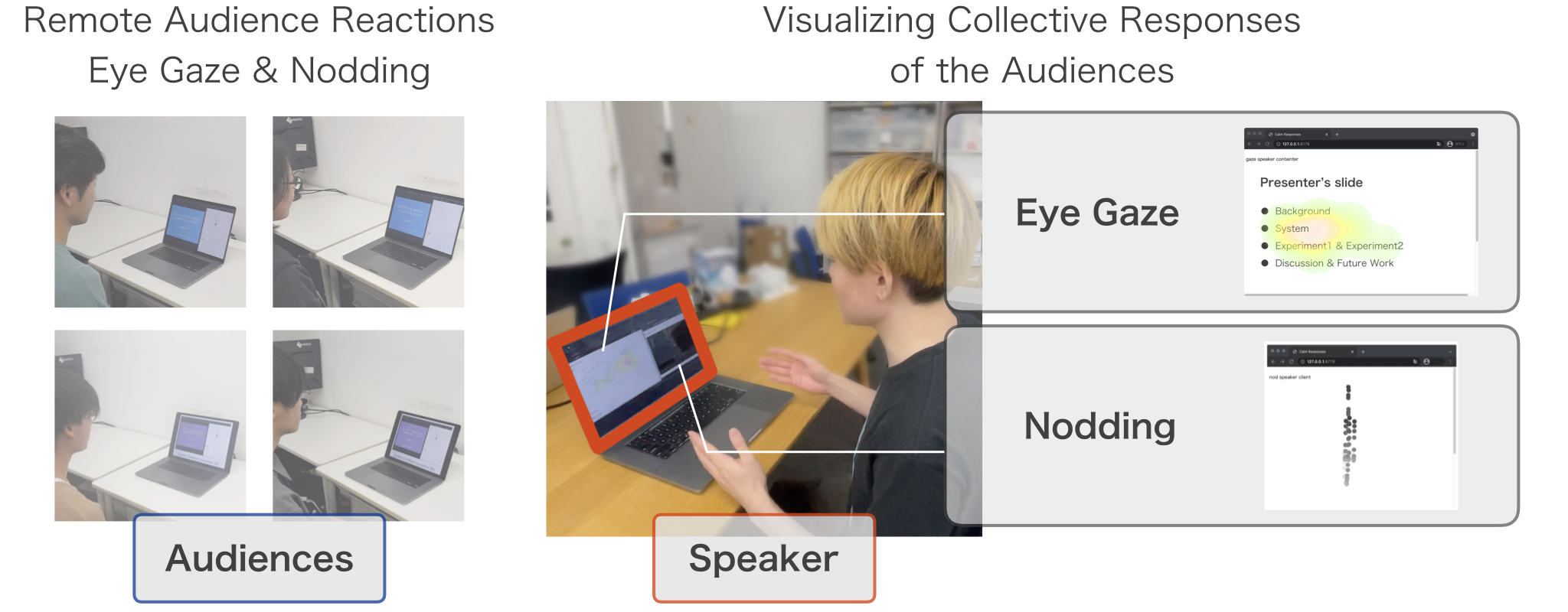}
\caption{The overview of the proposed system. The system obtains eye gaze and nod reactions of audiences with webcams (left) and collectively presents them to speakers in real time (right) during remote communication.}
\label{fig:teaser}
\end{teaserfigure}

\begin{abstract}
We propose a system displaying audience eye gaze and nod reactions for enhancing synchronous remote communication. Recently, we have had increasing opportunities to speak to others remotely. In contrast to offline situations, however, speakers often have difficulty observing audience reactions at once in remote communication, which makes them feel more anxious and less confident in their speeches. Recent studies have proposed methods of presenting various audience reactions to speakers. Since these methods require additional devices to measure audience reactions, they are not appropriate for practical situations. Moreover, these methods do not present overall audience reactions. In contrast, we design and develop CalmResponses, a browser-based system which measures audience eye gaze and nod reactions only with a built-in webcam and collectively presents them to speakers. The results of our two user studies indicated that the number of fillers in speaker's speech decreases when audiences' eye gaze is presented, and their self-rating score increases when audiences' nodding is presented. Moreover, comments from audiences suggested benefits of CalmResponses for them in terms of co-presence and privacy concerns.
\end{abstract}



\maketitle

\section{Introduction}
\label{sec:intro}

We frequently speak to others in distant locations using online communication tools. Some people give lectures to students remotely, and others present to groups of people in online meetings. One of the advantages of such online communication is that we are not constrained to physical space. We can access this communication style from anywhere as long as we connect to the Internet with laptops or smartphones. Due to the COVID-19 pandemic, there has been a soaring demand for online communication \cite{ting2020digital}, especially in education \cite{basilaia2020transition,mukhtar2020advantages} and work \cite{wang2021achieving} situation. As a result, many institutions have undergone a rapid transition to remote environments.

Despite its benefits and demand, audio- and video-based online communication has a crucial problem: it is difficult for speakers to see others' reactions compared to face-to-face communication. Existing teleconferencing tools, such as Skype, cannot effectively convey non-verbal cues \cite{lo2016skype}. This disadvantage is particularly remarkable when there are multiple audiences. For example, most students do not want to turn on their videos due to various reasons such as privacy concerns (e.g. appearance, background) as well as problems in their internet connection \cite{castelli2021students}. Even if audiences turn on these features, speakers cannot observe the reactions of the entire audience at a glance unlike offline situations due to the limited screen size or screen sharing \cite{kushalnagar2020teleconference}. When speakers cannot receive these non-verbal signals from audiences, they feel anxious and stressed during their speech \cite{bassett1973effects,macintyre1997effects}.

In response, several techniques have been proposed in the HCI literature to support speakers during online communication by transmitting audience reactions to them. One way to present audience reactions to speakers is for audience members to send texts or emoticons \cite{oomori2020preliminary,teevan2012displaying}. Although speakers can observe real-time reactions through this method, it forces audience members to manipulate laptops explicitly, making them feel bothersome. Hence, speakers' ability to obtain enough information depends on the active participation of the audiences. Moreover, texts or emoticons cannot necessarily reflect the real thoughts of their senders \cite{hassib2017heartchat,lo2008nonverbal}. Another way to communicate reactions is for audience members to share unconscious behaviors represented by non-verbal cues or physiological signals of audiences with speakers. This method can obtain audience reactions without burden.
Existing research has exploited these signals as audience reactions \cite{di2018unobtrusive,latulipe2011love,murali2021affectivespotlight,sun2019presenters}.

However, most of these systems require extra sensors in addition to laptops to obtain these reactions. Thus, these methods can only be used in offline controlled laboratory settings \cite{gashi2019using,hassib2017engagemeter,sugawa2021boiling,vatavu15}. Also, some systems acquire reactions obtrusively (e.g. attaching a sensor), which can negatively affect the communication experience for the audience. To calmly obtain entire audience reactions online, we require scalable systems that only use commonly available devices, such as laptops, and do not depend on any additional device. Furthermore, it is desirable for speakers to see overall audience reactions simultaneously as they can in offline situations. In other words, we need to design feedback that aggregates their reactions so that speakers can experience the communication atmosphere on a single screen.


In this study, we propose CalmResponses, a browser-based system that obtains the eye gaze and nod reactions of remote audiences using a webcam and collectively presents them to speakers. Through a long time of research \cite{carter2020best,darwin1998expression}, it is known that eye gaze and nodding are some of the most meaningful non-verbal cues in face-to-face communication. The same is true for online communication, and recent studies have shown that presenting eye gaze or nod reactions in real-time during conversation positively affects speakers \cite{chollet2015exploring,chollet2017perception,kubota2019speech}. CalmResponses presents these signals collectively, which enables presenters to see overall audience reactions simultaneously. In addition, since the proposed system does not require an additional device other than a webcam equipped with a laptop, it is easy to use in actual online settings.

We conducted two experiments and studied how collective audience reactions affect speakers through the objective and subjective evaluations of presentations, number of fillers, and qualitative comments about the presentation experiences. As a result, we found that CalmResponses positively affected not only the speakers but also the audiences. Specifically, the collective feedback of eye gaze reduced the number of fillers in speakers' speech and the collective feedback of nodding enhanced their self-rating scores. Moreover, the audiences found the experience positively, mentioning that it contributed to their feeling of co-presence as well as their low privacy invasiveness.

To summarize, our contributions are as follow:

\begin{itemize}
    \item We designed and developed a browser-based system that collectively presents real-time eye gaze and nod reactions of entire audience in remote communication settings to speakers using the built-in webcam.
    \item We conducted online experiments to study how the system affected the speakers' presentations and the result indicated that displaying eye gaze reactions reduced speakers' number of fillers, and displaying nodding increased self-rating scores.
    \item Based on the findings of the experiments, we discussed the potential benefits of collective visualization in terms of co-presence and privacy concerns for future audience sensing and feedback technologies.
\end{itemize}

\section{Related Work}
\label{sec:related_work}


\subsection{Audience Sensing and Feedback}

As mentioned in the introduction, one often-used method of presenting real-time audience reactions to speakers in online communication is to use texts or emoticons \cite{li2019live,lu2021streamsketch,oomori2020preliminary,teevan2012displaying,zhou2019magic}. For example, Teevan et al. \cite{teevan2012displaying} proposed a system that presents audience reactions, such as ``like'' or ``dislike'', in a slide during a presentation. Zhou et al. \cite{zhou2019magic} investigated the relationship in live streams between paid
gifting \footnote{paid gifting is a (virtual) gift or a donation that viewers can send to streamers.} and stimuli extracted from {\it danmaku}, moving comments on videos like bullets, and found that some variables such as number of comments positively affected paid gifting. Although these methods enable speakers to observe real-time audience reactions, they also force audiences to manipulate some devices explicitly such as pushing buttons. In other words, the success of these methods depends on active participation of the audience members.

In contrast, other studies have leveraged unconscious behaviors and states of audiences, such as non-verbal cues or physiological signals, and presented them to speakers \cite{di2018unobtrusive,hassib2017engagemeter,sugawa2021boiling,sung2021learners,vatavu15,yao2018visualizing}. For example, EngageMeter \cite{hassib2017engagemeter} allows presenters to observe audiences' engagement and workload through electroencephalography (EEG) signal. Yao et al. \cite{yao2018visualizing} developed a system that presents students' gaze movements to a teacher in online exercise lectures. They showed that this system can help the teacher to grasp students' attention and the progress of their tasks. Despite their efficacy, however, most of these methods require additional devices for sensing the reactions of the audience, such as Tobii \footnote{https://www.tobii.com} to estimate eye gaze positions. Moreover, these systems often use obtrusive data collection methods, negatively affecting the audience experiences. It is important to use simple interaction technologies that do not interfere with audiences experiences \cite{barkhuus2008engaging}.

To obtain real-time reactions of multiple audiences in online situations unobtrusively, we require a scalable system that only uses commonly available devices, such as laptops. In this sense, AffectiveSpotlight \cite{murali2021affectivespotlight} is a suitable system to share audience reactions with speakers because it estimates facial responses and head gestures of audiences with a webcam. However, it presents only one audience with the highest engagement simultaneously and it is impossible for speakers to grasp how the others feel because it is hard to deny that while one audience is engaging, the others are not during presentations. Given the assumed scenarios such as education as we mentioned in the introduction, it would be better for speakers to observe the collective audience reactions simultaneously. Thus, in this paper, we develop a system that can unobtrusively sense audience reactions and visually present them to speakers in a collected manner.

\subsection{Non-verbal Behavior Analysis  in Communication}
\label{sec:related_work_nonverbal}

Eye gaze and nodding has been subjects of research in human communication for a long time \cite{darwin1998expression,mcclave2000linguistic,wagner2014gesture}. On one hand, eye gaze has been utilized as a cue to estimate various internal states since it has been shown to be correlated with human cognitive processes closely \cite{carter2020best}. For example, we can infer one's interest or attention \cite{cherubini2008deixis,DBLP:conf/chi/ArakawaY19,hutt2021breaking}, one's linguistic abilities \cite{fujii2019subme}, the type of books one is reading \cite{kunze2013know}, and even what picture one is recalling \cite{wang2019mental} based on eye gaze movements. On the other hand, nodding is also one of the main methods of conveying information in communication, as we can see from a finding that over 80\% of listeners' head movements are nodding \cite{wlodarczak2012listener}. In face-to-face communication, nodding has various meanings and functions \cite{hale2020you,poggi2010types}. For example, listeners' nods indicate confirmation, agreement, and even disagreement \cite{poggi2010types}.

These days, researchers study the use of these modalities in online communication. For eye gaze, prior research has explored the effect of sharing eye gaze movements with remote users in online communication \cite{d2016gazed,d2017improving,sung2021learners,vertegaal1999gaze,vspakov2019two,yao2018visualizing}. In these cases, users share their gaze positions on their screens, which helps to disambiguate what the speaker is referring to \cite{d2017improving}. GAZE groupware system \cite{vertegaal1999gaze} is a system that conveys gaze directions in multiparty mediated communication. In this system, users can easily grasp who is talking to whom and about what. Other research \cite{vspakov2019two,yao2018visualizing} visualized eye gaze positions from multiple students in remote classes or exercises. Similarly, nodding has also been leveraged in computer-mediated communication \cite{chollet2015exploring,gupta2019blink,kubota2019speech,maloney2020talking,watanabe2004cosmic}. For example, Kubota et al. \cite{kubota2019speech} proposed a system that presents nodding images related to spoken words in online communication. Chollet et al. also developed a system leveraging virtual audiences for presentation training \cite{chollet2015exploring}. This virtual audience responds with non-verbal cues, including nodding during presentations. Likewise, Maloney et al. \cite{maloney2020talking} investigated non-verbal communication in social virtual reality and found that nodding is a naturally used interaction.

While previous research has proposed systems collectively visualizing engagement of multiple audiences through facial expressions \cite{sun2019presenters} and heart rate \cite{sugawa2021boiling}, little is known about how we design the collective visualization from multiple audiences' eye gaze and nodding reactions. In the present study, we propose a system that presents audience eye gaze and nod reactions collectively to speakers. The system estimates and displays the users' eye gaze and head movements (specifically nod reactions). In the next section, we introduce how we estimate and display these reactions without relying on external devices other than conventional laptops.

\section{Proposed System: CalmResponses}

\subsection{System Overview}
\label{sec:system_overview}

We propose CalmResponses, which displays audience eye gaze positions and head movements collectively to speakers for supporting them to grasp audience reactions in real time. Based on the existing research and issues as we discussed in Section \ref{sec:related_work}, we need to develop the system appropriate for real-time, remote, and multiple-audience situations. Specifically, we aim to meet the following two requirements.

\paragraph*{{\bf Do not need any additional device to unobtrusively estimate reactions}}

Although we can obtain richer information with additional devices \cite{hassib2017engagemeter}, it increases the cost to utilize in actual online settings. To estimate and display entire audience reactions while keeping availability, we implemented the system which functions only with a built-in webcam and a web browser.

\paragraph*{{\bf Display eye gaze positions and head movements of multiple audiences collectively}}

To present reactions of multiple audiences, we need to aggregate reactions on a single screen. The aggregation of reactions from multiple users has also been proposed in \cite{sun2019presenters} leveraging facial expressions. We designed and developed collective feedback method for eye gaze and nodding.

\vspace{0.2in}

CalmResponses satisfies the above requirements, implemented as a browser-based application using HTML, CSS, and JavaScript. The system consists of three sub-systems: audience client, speaker client, and server. The audience client obtains and sends audiences' eye gaze or head movements through a built-in webcam, and the speaker client receives and displays feedback in real time. We adopted WebSocket to connect the server with both the speaker and audience clients. We deployed the system on Heroku. We provide the source code \footnote{https://github.com/kiyosumaeda/CalmResponses} and the link \footnote{https://calmresponses.herokuapp.com/} to try the system. Users can take this application together with other video conferencing tools such as Zoom as long as they open the browser interface. In the next section, we elaborate on the audience client and speaker client respectively, centering on how they estimate and display eye gaze positions or head movements of audiences.

\subsection{Collective Eye Gaze Reactions}

\subsection*{Sensing}
\label{sec:system_gaze_estimate}

To estimate one's eye gaze position on the screen, we used WebGazer.js \cite{papoutsaki2016webgazer}, a javascript library to infer eye gaze positions based on webcams equipped with laptops in real time. It has shown as a reliable tool for eye tracking due to its low errors. Since this library requires only a built-in webcam and the browser, it is suitable for our purpose, as discussed in Section \ref{sec:system_overview}.

Users need to conduct calibration before using the system. We introduced the calibration process at the beginning of the usage. We followed the original research \cite{papoutsaki2016webgazer} to implement the calibration process in the audience client.

Since raw gaze data are not stable even in fixation due to the involuntary movements of eyes \cite{yarbus1967eye}, we need to smooth the data. While there are various methods to smooth them \cite{kumar2008improving}, we concisely smooth them by calculating the mean values of gaze positions for several frames. Thus, the audience client sends relative eye gaze positions smoothed on the screen to the speaker client through a server. Next, we show how we display gaze positions in the speaker client.

\subsection*{Feedback Design}

\begin{figure*}[t]
\centering
\includegraphics[width=15cm]{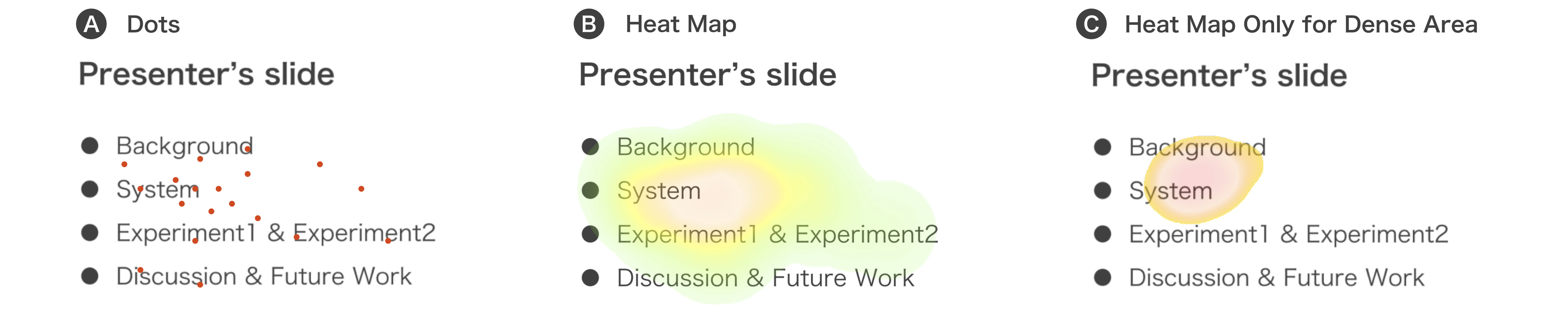}
\caption{Various candidate ways to display audience eye gaze movements: Dots (A), Heat Map (B), and Heat Map for only dense area (C)}
\label{fig:eye_gaze_variation}
\end{figure*}

In this section, we describe how we present eye gaze movements of entire audience collectively. Figure \ref{fig:eye_gaze_variation} shows the three ways to visualize eye gaze movements. 
Specifically, we considered three ways for presenting gaze collectively: Dots, Heat Map, and Heat Map only for the dense area.

\subsubsection*{Dots}

Dots are the most common approach to visualize gaze data \cite{d2018eye}. As described in Figure \ref{fig:eye_gaze_variation}A, red dots on eye gaze positions of each user were presented. In detail, upon receiving data of relative gaze position from the audience client, the speaker client calculates an absolute gaze position. Although this approach is a primitive way to visualize gaze positions, one of the problems is that this dot visualization can be distracting for speakers \cite{d2016gazed}.

\subsubsection*{Heat Map}

Figure \ref{fig:eye_gaze_variation}B shows the heat map visualization of eye gaze data. Previous work has adopted this heat map visualization \cite{d2018eye,newn2017evaluating}. The color indicates the density of the audience gaze (e.g. red areas are where the audience is mostly looking at). The heat map is translucent so that users can see original contents under the visualization without being distracted. This visualization can help speakers to understand audience interests and attitudes intuitively. For example, from red areas, speakers can see which part of the content the audiences are interested in. Likewise, when the red area is exactly where the speakers are referring, they can see that audiences pay attention to their speeches. Conversely, when there is no dense area, they can infer that audiences are likely to be distracted. In this way, heat map visualization helps speakers understand the audience reactions based on their gaze movements. Heat map is also robust to noise that derives from accuracy because heat map represent ranges (not the exact position) \cite{d2018eye}, which is suitable for actual remote situations.

\subsubsection*{Heat Map Only for Dense Area}

Another way to display gaze information is to draw only dense areas of heat map as in Figure \ref{fig:eye_gaze_variation}C. This corresponds to Shared Area \cite{d2018eye} where multiple users look at simultaneously. This visualization allows users to focus more on where many audiences are looking. On the other hand, this visualization is hard to reflect the minority's interests. Moreover, when audience eye gaze positions are distributed within the screen, it does not represent overall audience reactions.

\vspace{10px}
Considering the advantages of Heat Map as real-time feedback for speakers, we adopted Heat Map visualization for CalmResponses.

\subsection{Collective Nod Reactions}

\subsection*{Sensing}

We estimate head movements based on facial landmarks. We use clmtrackr\footnote{https://www.auduno.com/clmtrackr/}, a javascript library to detect facial landmarks in videos or images. Since this library requires only a built-in webcam and a browser, it is suitable for online settings as we discussed in Section \ref{sec:system_overview}.

While there are various methods to estimate head movements \cite{bousmalis2013towards}, we estimate them only from one landmark corresponding to the nose tip position. The rationale of this method is as follows: on one hand, when we nod, we rotate our heads around the inter-aural axis \cite{wagner2014gesture}. In this condition, the nose tip position also moves up and down. On the other hand, when we shake our heads, we move our heads laterally. In this condition, the nose tip position also moves left and right.  Therefore, we can estimate most head movements only from the nose tip position. Although this method cannot distinguish still head from other head movements, such as head rotation around the naso-occipital axis which expresses disbelief, we did not consider those because we mainly focused on nodding as we discussed in Section \ref{sec:related_work_nonverbal}.

\subsection*{Feedback Design}

\begin{figure*}[h]
\centering
\includegraphics[width=14cm]{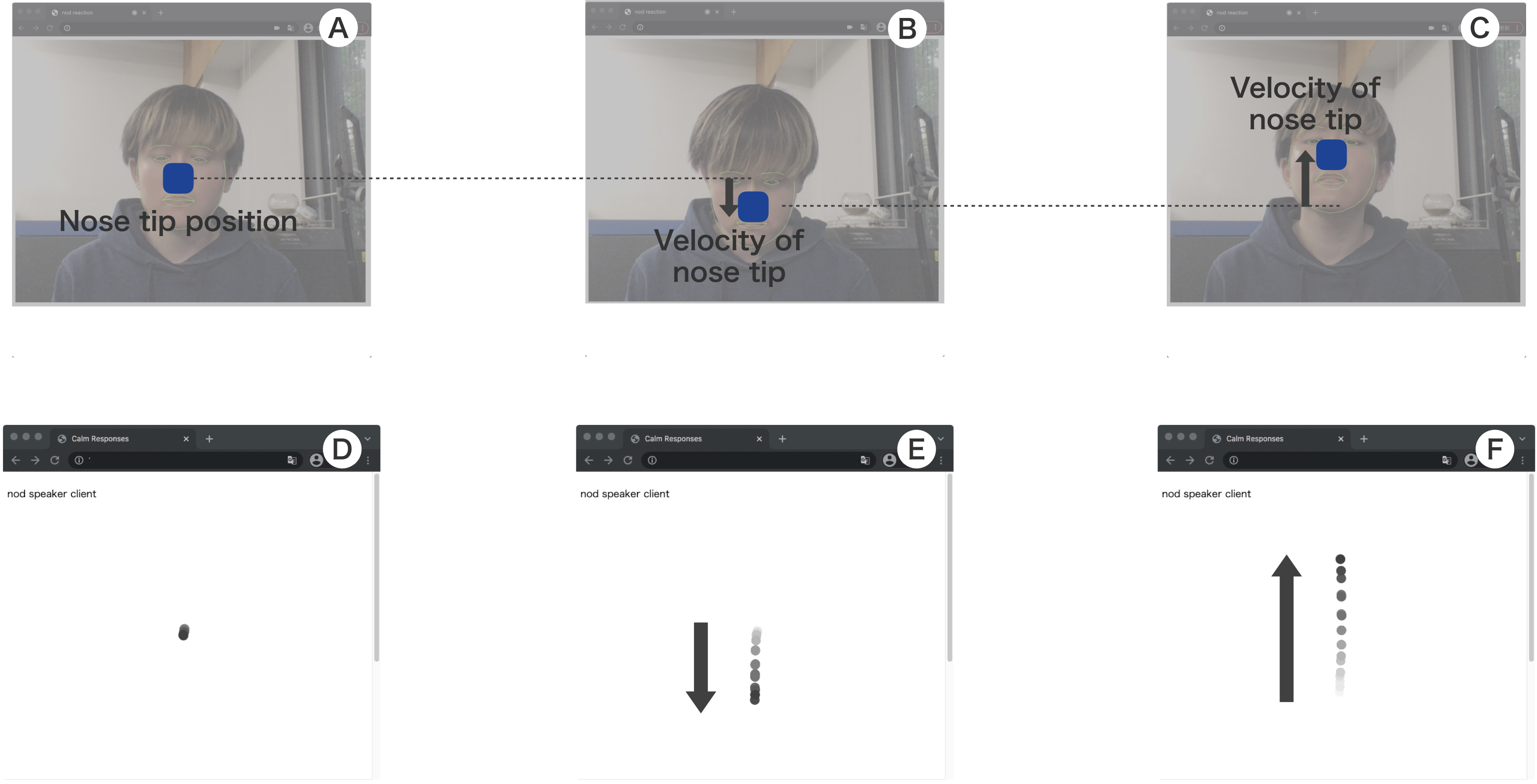}
\caption{Illustration of how the system estimates and visualizes a user's head movement. When the user faces the front (A), the cursors' trails are still in the browser's center (D). Next, when the user shakes their head down (B), the nose tip position's velocity vector is downward direction. Then, the trails of cursors extend from center to bottom (E). Finally, when the user shakes their head up (C), the nose tip position's velocity vector is in the upward direction. Thus, the trails extend from bottom to top (F).}
\label{fig:nod_display}
\end{figure*}

We use trails of cursors to visualize head movement. Compared to the gaze visualization, there is little research regarding how to visualize head movements. We adopted cursors to visualize them more intuitively than other methods such as numbers (e.g. how many people are nodding). Figure \ref{fig:nod_display} shows an example of how the system estimates and visualizes a user's head movement. The trails of cursors move based on the nose tip position's velocity vector. For example, when the user shakes their head down (Figure \ref{fig:nod_display}B), the nose tip position's velocity vector is downward direction. In this case, the trails of cursors extend from center to bottom (Figure \ref{fig:nod_display}E).

\begin{figure*}[t]
\centering
\includegraphics[width=6.5cm]{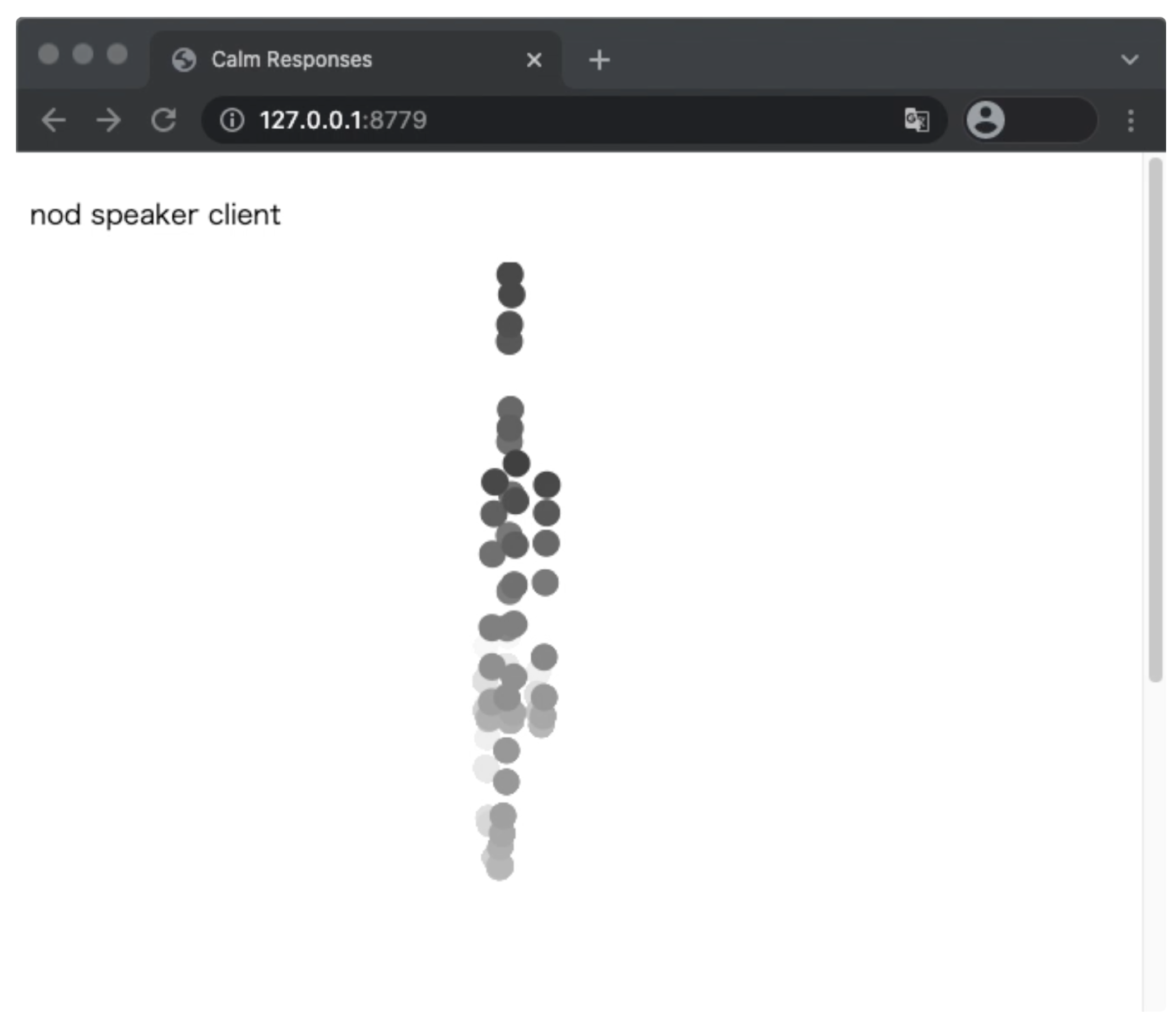}
\caption{Collective visualization of head movements. We superimpose all audience's trails of cursors with some offsets.}
\label{fig:collective_display}
\end{figure*}

Next, we describe the way we present the head movements of multiple users. We superimpose all audiences' trails with some horizontal offsets. Figure \ref{fig:collective_display} shows an example visualization of trails from three audiences. Since we overlay the multiple users' trails together, similar head movements are emphasized by overlapping many cursors. For example, when speakers observe the system and notice that cursors are extending vertically, they can infer that many audiences are nodding. This is the benefit of collectively displaying audience reactions as we can observe clearer vertical feedback in Figure \ref{fig:collective_display} than in Figure \ref{fig:nod_display}. While speakers can see the gaze feedback that is visualized on their slide directly (See Figure \ref{fig:eye_gaze_variation}), the nod feedback was presented alongside their slide (See Figure \ref{fig:exp2_screen}).


We calculate the velocity of the nose tip landmark about every 30 ms and send that velocity to the speaker client through the server. When the speaker client receives new data, it updates the position of each cursor. We can change the amplitude of horizontal and vertical movements so that the system can emphasize nodding.

\section{Experiment Overview}
\label{sec:experiment}

Up to this point, we introduced our proposed system estimating and displaying collective reactions of multiple audiences.
We conducted two experiments to evaluate the efficacy of the system displaying collective eye gaze and nodding reactions in online communication.

\subsection{Design}
\label{sec:experiment_purpose}

We conducted two experiments that replicated online communication situations. Specifically, we asked participants to have presentations to audiences about given topics for a few minutes on Zoom, a common video conferencing platform. We employed a within-participants design comparing a treatment condition using CalmResponses with a baseline condition. Namely, there are three conditions: {\it condition B}, {\it condition CR-E}, and {\it condition CR-N}.

\begin{itemize}
    \item {\it Condition B} : Baseline condition. In this condition, the audience turned off their video and audio. In other words, speakers could not see any reaction from the audience. This condition is based on the finding from existing research that most students tend to turn off their video and audio during online classes \cite{castelli2021students}.
    \item {\it Condition CR-E}: System (CalmResponses) condition with eye gaze reactions. Although speakers could not see any video and audio from the audience, same as in {\it condition B}, they could see the collective eye gaze positions of the audience through the speaker client of the system (see Section \ref{sec:system_overview}).
    \item {\it Condition CR-N} : System (CalmResponses) condition with nod reactions. In this condition, although speakers could not see any video and audio from the audience, same as in {\it condition B}, they could see the collective head movements of the audience through the speaker client of the system.
\end{itemize}

\noindent
In the first experiment, we compared {\it condition B} and {\it condition CR-E}; and in the second experiment, we compared {\it condition B} and {\it condition CR-N}.
In each experiment, we performed three evaluations of the system: objective and subjective evaluation of the presentations, number of fillers, and qualitative comments.
These are widely used measures in existing works to evaluate speech quality and indirectly estimate speakers' states such as anxiety \cite{murali2021affectivespotlight,trinh2017robocop}.

\subsection{Measure}
\label{sec:experiment_measure}

\paragraph*{{\bf Objective and Subjective Evaluation of the Presentations}}

We used a questionnaire to evaluate the speaker's speech quality based on RoboCOP \cite{trinh2017robocop} after presentation. Speakers self-evaluated their speech quality, and the audience evaluated the speakers' speech quality. The questionnaire contained six items, and each item was rated from one to seven (1 = Not At All, 7 = Very Much). We calculated the sum of these scores for every speaker (note that we added reversed scores for the third question). The questionnaire for the speakers contained the following questions:

\begin{itemize}
    \item How engaging was your presentation?
    \item How understandable was your presentation?
    \item How nervous were you during your presentation?
    \item How exciting was your presentation?
    \item How entertaining was your presentation?
    \item How competent were you during your presentation?
\end{itemize}

\noindent
Note that the above questionnaire was used for the speakers. When asking the audience, we modified each item to be suitable for them. For example, ``How engaging was your presentation?'' was changed into ``How engaging was the presenter's presentation?''
We hypothesized that the system would help participants increase their evaluation scores.

\paragraph*{{\bf Number of Fillers}}

In addition to the questionnaire evaluation, we evaluated the presentations using automatic speech analysis. We examined the number of verbal fillers used from the recorded presentation data to estimate speech performance. Since it is known that fillers are negatively correlated with speakers' performance or anxiety \cite{chen2014towards,chollet2015exploring,goberman2011acoustic,scherer2012audiovisual}, we anticipated that the number of fillers per minute would decrease in {\it condition CR-E} and {\it condition CR-N} compared to {\it condition B}.

\paragraph*{{\bf Qualitative Comments}}
\label{purpose_qualitative_analysis}

While the above two measures were effective, we also evaluated the system qualitatively. After the experiments, we asked the speakers (1) how and why they changed their self-rating scores compared to {\it condition B}, (2) how they perceived and interpreted the presented feedback, and (3) what the pros and cons of the system were. We also asked the audiences about their overall experience of the experiments. Specifically, we asked them how they perceived the system and how they reacted to the speakers.

\subsection{Participants}

\begin{table*}[t]
\caption{The groups of participants in the experiments. There are six groups of five to seven participants. We use SE1-SE5 / AE1-AE19 to denote speakers / audiences in experiment 1 and SN1-SN6 / AN1-AN8 to denote speakers / audiences in experiment 2.}
\begin{tabular}{cccc}
group  & conditions & number of speakers & number of audiences \\ \hline
group1 & B \& CR-E (experiment 1) & 1 (SE1) & 4 (AE1 - AE4) \\
group2 & B \& CR-E (experiment 1) & 2 (SE2, SE3) & 5 (AE5 - AE9) \\
group3 & B \& CR-E (experiment 1) & 1 (SE4) & 5 (AE10 - AE14) \\
group4 & B \& CR-E (experiment 1) & 1 (SE5) & 5 (AE15 - AE19) \\
group5 & B \& CR-N (experiment 2) & 3 (SN1 - SN3) & 4 (AN1 - AN4) \\
group6 & B \& CR-N (experiment 2) & 3 (SN4 - SN6) & 4 (AN5 - AN8)
\end{tabular}
\label{tab:experiment_participant}
\end{table*}

We recruited 38 participants (nine females, 29 males, ages 19-44, mean 24.7).  All participants were Japanese or Chinese, and all could fluently speak and understand Japanese. We randomly split them into six groups of five to seven participants as described in Table \ref{tab:experiment_participant}. They did not know each other at the time of the experiments. In each group, we randomly assigned one to three participants to be speakers, while the other members (four to five) served as an audience. The experiment was conducted online using Zoom, an online conferencing tool. In total, five speakers gave presentations in Experiment 1 (comparing {\it condition B} and {\it condition CR-E}), and six speakers in Experiment 2 (comparing {\it condition B} and {\it condition CR-N}). We use SE and SN to denote speaker participants in experiment 1 and experiment 2 respectively, while we use AE and AN to denote audience participants in experiment 1 and experiment 2 respectively.

\subsection{Procedure}
\label{sec:exp_procedure}

\begin{figure*}[h]
\centering
\includegraphics[width=15cm]{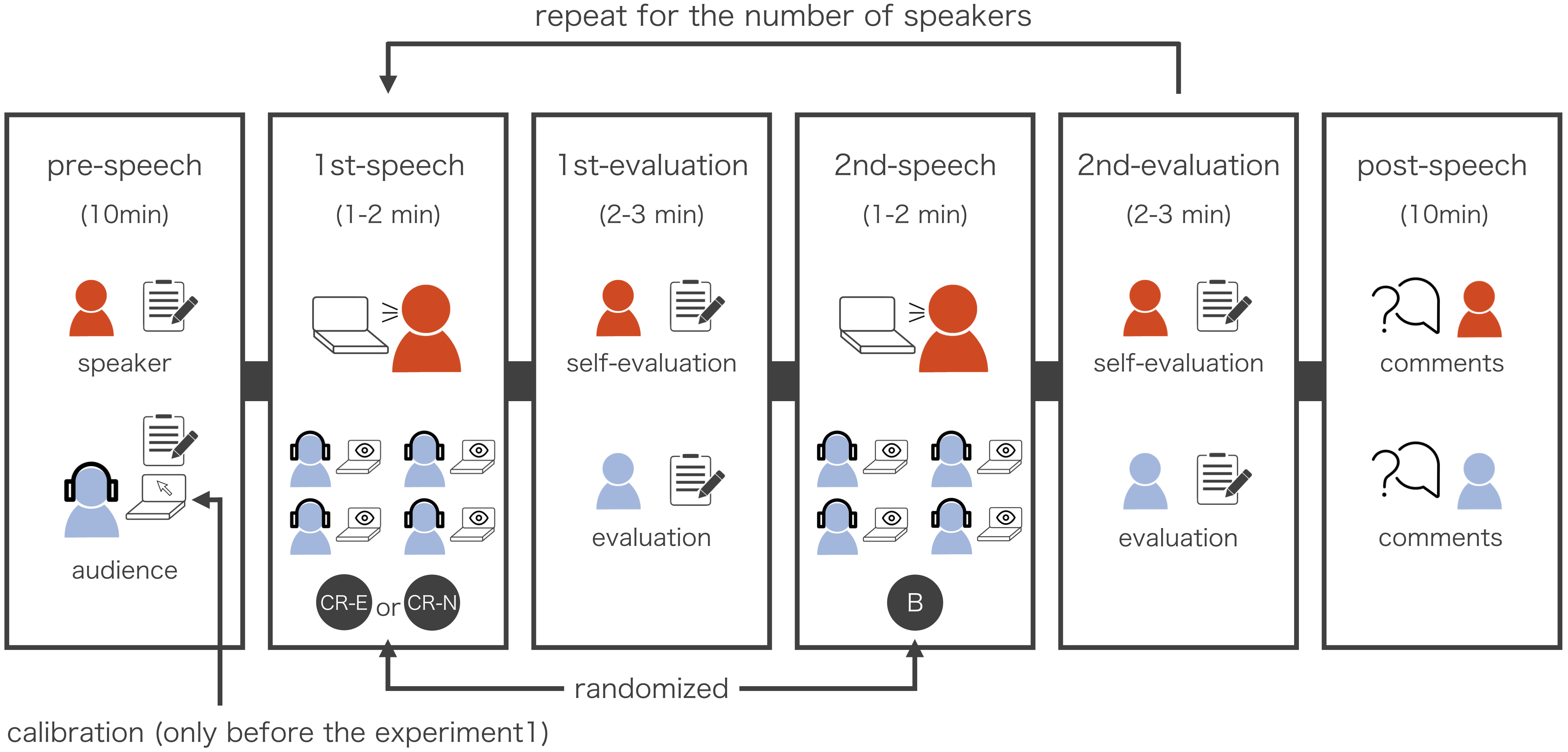}
\caption{The overview of the experiment 1 and experiment 2.}
\label{fig:experiment_setting}
\end{figure*}

The overview of the two experiments is presented in Figure \ref{fig:experiment_setting}. The experiments contained four types of sessions: a pre-speech session, speech sessions (the first and the second), evaluation sessions (the first and the second), and a post-speech session. In the pre-speech session, all participants answered a questionnaire about their demographics. After finishing the questionnaire, the audiences accessed CalmResponses using Google Chrome, and an experimenter briefly explained how to use the system. Audiences were asked to put a webcam and a display in front of them.

In the speech-session, the speakers were asked to give presentations in Japanese twice to the audiences on given topics.
They talked about different topics for the two presentations, and the topics were informed in advance.
Although they could not read manuscripts, they were allowed to prepare their thoughts before their presentations.
Each presentation lasted for approximately two minutes.
While this was a short time for a presentation, speakers did not need to make an effort to prepare, which could led to the natural speech.
Before each presentation, speakers filled out a questionnaire about the State Anxiety \cite{spielberger1989state} to check their anxiety level for the presentation. One presentation was given under {\it condition B}, while the other was given under {\it condition CR-E} (experiment 1) or {\it condition CR-N} (experiment 2). The order of conditions was randomized. The total time of the speech and evaluation sessions was about 10-30 minutes ((2-min speeches + 3-min evaluation periods) * (1 to 3)-presenter * 2-condition). After a speaker finished their presentation, the speaker and the audience answered a questionnaire on the presentation quality for about a few minutes (evaluation session). The speakers could not observe evaluations by the audiences and vice versa. Each speaker did not listen to the other speakers' presentations. A series of speech and evaluation sessions were repeated for each speaker. Finally, in the post-speech session, we collected their comments regarding the usability of the system, as we mentioned in Section \ref{sec:experiment_measure}. The audio of the presentations was recorded for analyzing the number of fillers.

\section{Experiment 1: Displaying Eye Gaze Reactions}
\label{sec:experiment1_title}

In this section, we explain the first experiment, which explored the use of collective eye gaze reactions of multiple audiences in online communication.

\subsection{Detailed Procedure}

\begin{figure}[h]
\centering
\includegraphics[width=8cm]{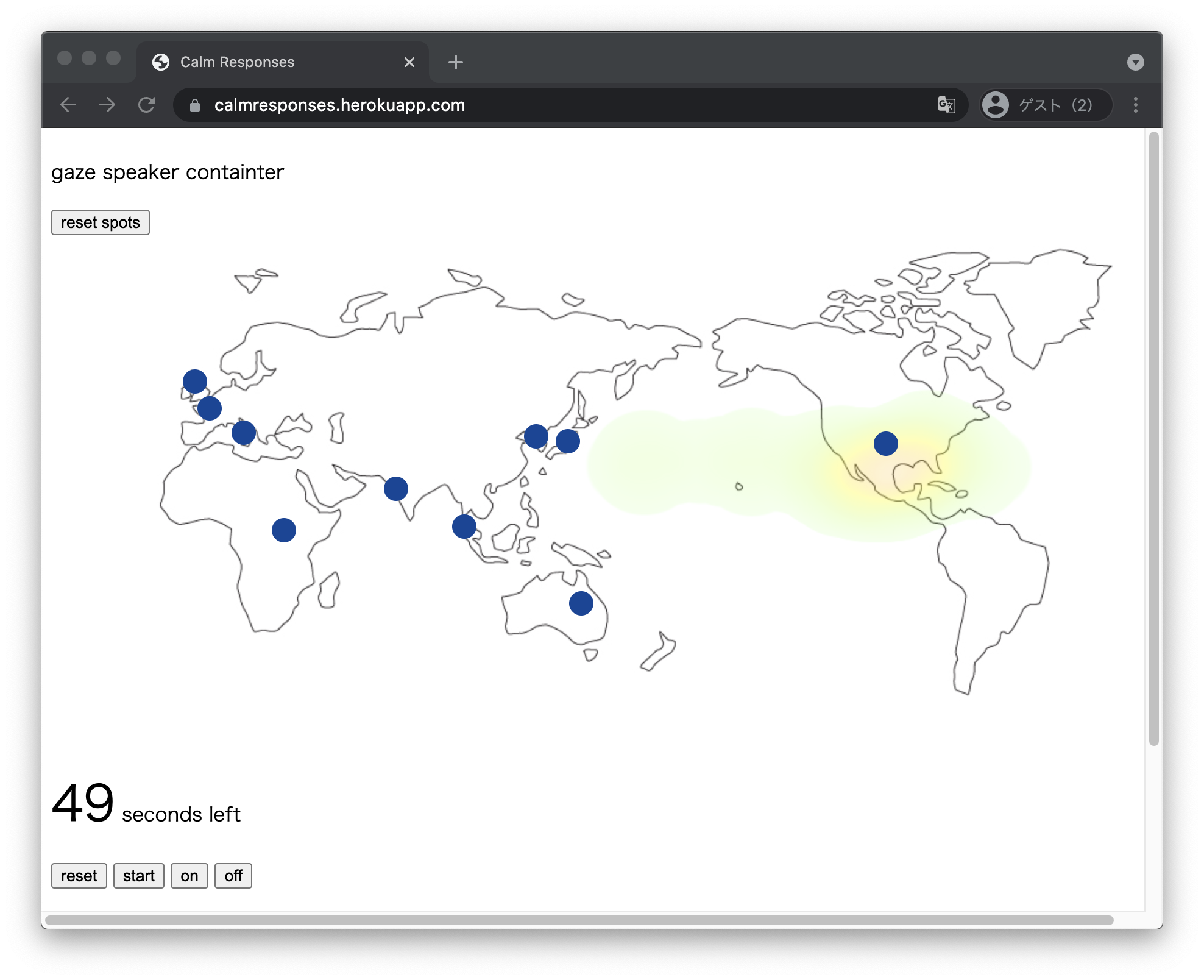}
\caption{View of the speakers in {\it condition CR-E}, in which the audience eye gaze movements were visualized as a heat map. In both {\it CR-E and B conditions}, the speakers were asked to give presentations while seeing a map (Japan or the world map). They could place blue markers in advance where they would explain in their speeches. These markers were also visible on the audience clients.}
\label{fig:experiment1_speaker_spots}
\end{figure}

After audiences filled out the questionnaire in the pre-speech session, they started to calibrate their gaze following an experimenter's instruction as we described in Section \ref{sec:system_gaze_estimate}. The topic in both {\it condition CR-E and B} was: ``Please explain experiences in places where you've been before.'' The speakers could choose choose places either in Japan or in the world. They gave their presentations while seeing a map of Japan or the world. They talked about different places in the two presentations. The audiences also saw the same map as speakers' during the presentation through the audience client. Figure \ref{fig:experiment1_speaker_spots} presents a speaker's browser screen. A heat map was visualized in {\it condition CR-E}. On the other hand, speakers could see only a map and the remaining time in {\it condition B}. The speakers were allowed to place blue markers on a map in advance where to explain in their speech. These markers were also visible on the audience clients.


\subsection{Result}

\begin{figure*}[t]
\centering
\includegraphics[width=15cm]{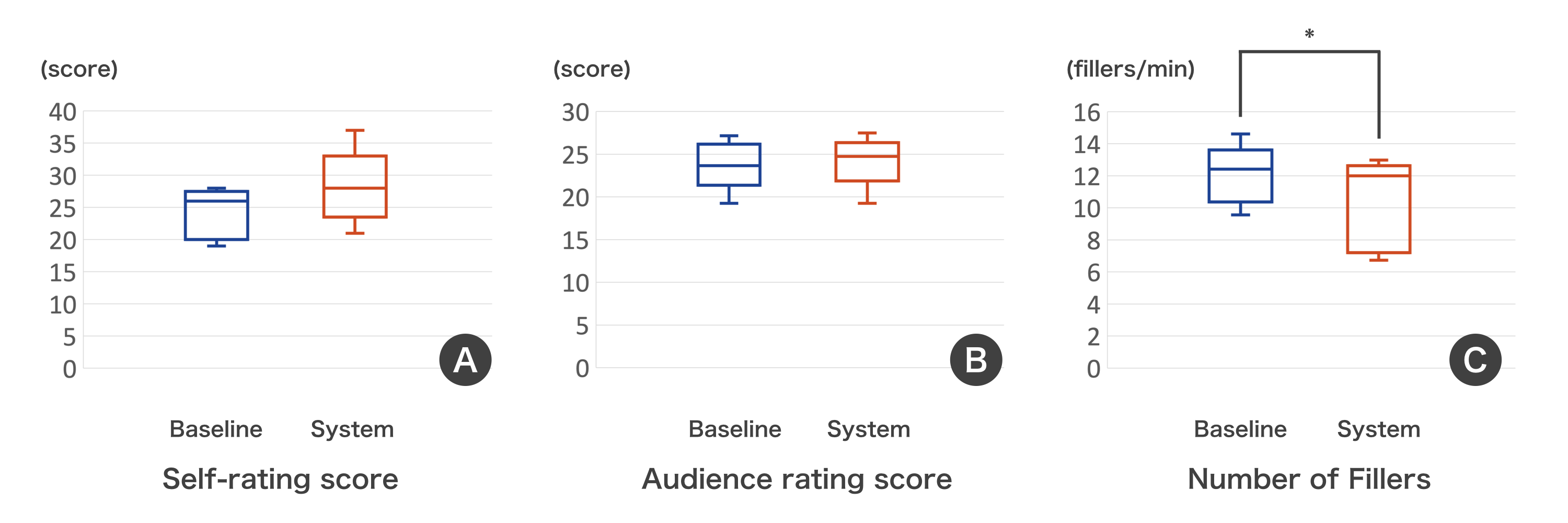}
\caption{Self-rating score (A), Audience rating score (B), Number of fillers per minute (C). p < .05 is marked as *.}
\label{fig:experiment1_rating}
\end{figure*}

\subsection*{How Did the Speakers Change Subjectively?}

The result of the paired t-test showed no significant difference between the baseline and system conditions on the State Anxiety (p = .36) which was measured right before the presentation (See Section \ref{sec:exp_procedure}). Although the average self-rating scores were higher in {\it condition CR-E}, there was no significant difference between the two conditions (p = .085) as denoted in Figure \ref{fig:experiment1_rating}A. We further analyzed the effect of CalmResponses on the speakers via collected comments. Three out of five speakers raised their self-evaluation scores. We found three strengths of the system through feedback from these speakers. First, they mentioned that they felt a sense of relief in using the system: {\it ``I felt more relieved because I could figure out whether the audiences understood my speech''} [SE1]. {\it ``heat map helped me feel at ease''} [SE2]. Second, the speakers mentioned that the system helped them feel more aware of the audience: {\it ``I thought audiences were listening to my speech''} [SE1]. {\it ``I became more aware of the presence of the audience''} [SE2]. {\it ``Since I found that eye gaze positions moved every time I changed the location, I could confirm that the audience did listen to my presentation''} [SE4]. Finally, one speaker reported feeling more confident about his speech: {\it ``I spoke with confidence, especially because I could tell if the audience understood me or not when I mentioned the names of places''} [SE4]. While all speakers interpreted eye gaze positions as audience interest or attention, it is interesting to note that some speakers also interpreted eye gaze as the audience understandings. These findings (feeling relieved, awareness of the audience, and confidence) are aligned with previous works \cite{murali2021affectivespotlight,parmar2020making,trinh2017robocop}. Thus, it is implied that the system could positively affect speakers.


There were two speakers whose ratings did not increase due to the introduction of CalmResponses. One speaker felt that the audience did not necessarily see where the speakers were talking about: {\it ``Although I felt relieved to know that audiences paid attention to me, at the same time, I felt a bit anxious because they sometimes saw where I was not explaining''} [SE3]. Although this was partly because of the inaccuracy of eye tracking, it is true that the audiences were not always paying attention to the same point that the speaker was talking about. The other participant pointed out that he did not see the visualization much: {\it ``Since I was concentrating on remembering my experience, honestly, I didn't see the display so often''} [SE5]. This participant also argued that the system would be more beneficial when using more materials for the presentations. In fact, some audience members mentioned the same point: {\it ``I thought that my eye gaze movements would changed more dynamically in slide-based lectures or presentations''} [AE5]. Therefore, we need to consider and compare various situations and explore more suitable ones for the visualization of eye gaze movements.

\subsection*{How Did the Speakers Change Objectively?}

Audience rating scores were not significantly different between the two conditions (p = .06) as described in Figure \ref{fig:experiment1_rating}B. We also found some comments from the audiences that supported this result: {\it ``I couldn't notice the difference between the two speeches''} [AE9]. In contrast, in terms of the number of fillers, we did find a significant difference (p = .04) as described in Figure \ref{fig:experiment1_rating}C. Since the number of fillers in speech is negatively correlated to with speakers' performance or anxiety \cite{goberman2011acoustic}, this result was incompatible with the result of audience rating scores. To understand more objective speakers' change, it will be a promising approach to use other measures such as speech evaluation by experts \cite{chollet2015exploring}.

\subsection*{How Did CalmResponses Affect the Audience?}

We asked the 19 audiences how they perceived the system. First, many of the audiences mentioned that it was easy to use the system (N=10): {\it ``I didn't need any complex procedure''} [AE7]. {\it ``Since all I needed was a laptop with a webcam, it was easy to introduce''} [AE8]. In contrast, two audiences had trouble with their laptops freezing in the middle of the {\it condition CR-E} (AE3, AE14), and a few audience members pointed out that it was inconvenient to use the system in addition to Zoom: {\it ``One of the disadvantages was that I had to open the system through Google Chrome other than Zoom''} [AE17]. We expect that CalmResponses could be integrated into existing teleconferencing tools to address this usability issue.

We also asked audience members where they were looking during presentations. All audience members reported they looked at the locations mentioned by the speakers. However, the system made a few audiences feel nervous (N=3): {\it ``I think it was weird for speakers both when I was gazing at a single point and when I changed points so frequently''} [AE4]. {\it ``I felt cramped a little because speakers easily found out whether we were listening or not''} [AE11]. These comments imply that we need to consider their privacy. Still, we believe our collective visualization using a heat map could mitigate the issue since it does not identify individual gaze movements.

Interestingly, on the other hand, some audiences claimed that they were encouraged by the system to actively participate in the presentations (N=5): {\it ``I felt that I needed to be a good listener''} [AE2]. {\it ``By being aware that my gaze positions were seen by speakers, I could focus more on listening to the speech''} [AE17]. In terms of online learning, it is difficult for learners to maintain their concentration, and several works have sought to address this problem \cite{arakawa2021mindless,hutt2021breaking}. Our system can provide an alternative approach to the issue, although we need to balance users' feelings between active participation and nervousness as we mentioned in the previous paragraph.

Finally, we asked the audience members whether they wanted to see other audience members' eye gaze positions. Most of them answered positively about this question (N=15): {\it ``I'm curious about what other audiences are interested in while listening to the presentations''} [AE3]. {\it ``When I knew where others were looking at, I could notice the difference and similarity of the reactions between others and me''} [AE16]. In addition, they mentioned that they wanted to watch their own eye gaze positions as well. This point is consistent with a previous research that has emphasized the importance of providing audiences with visual feedback on their own actions \cite{barkhuus2008engaging}. In response, we modified the experiment to share the audience reactions not only with the speakers but also with the audience participants when we conducted the second experiment.

\section{Experiment 2: Displaying Nod Reactions}

In this section, we explain the second experiment, which explored the use of collective nod reactions of multiple audiences in online communication.

\subsection{Detailed Procedure}

\begin{figure}[h]
\centering
\includegraphics[width=8cm]{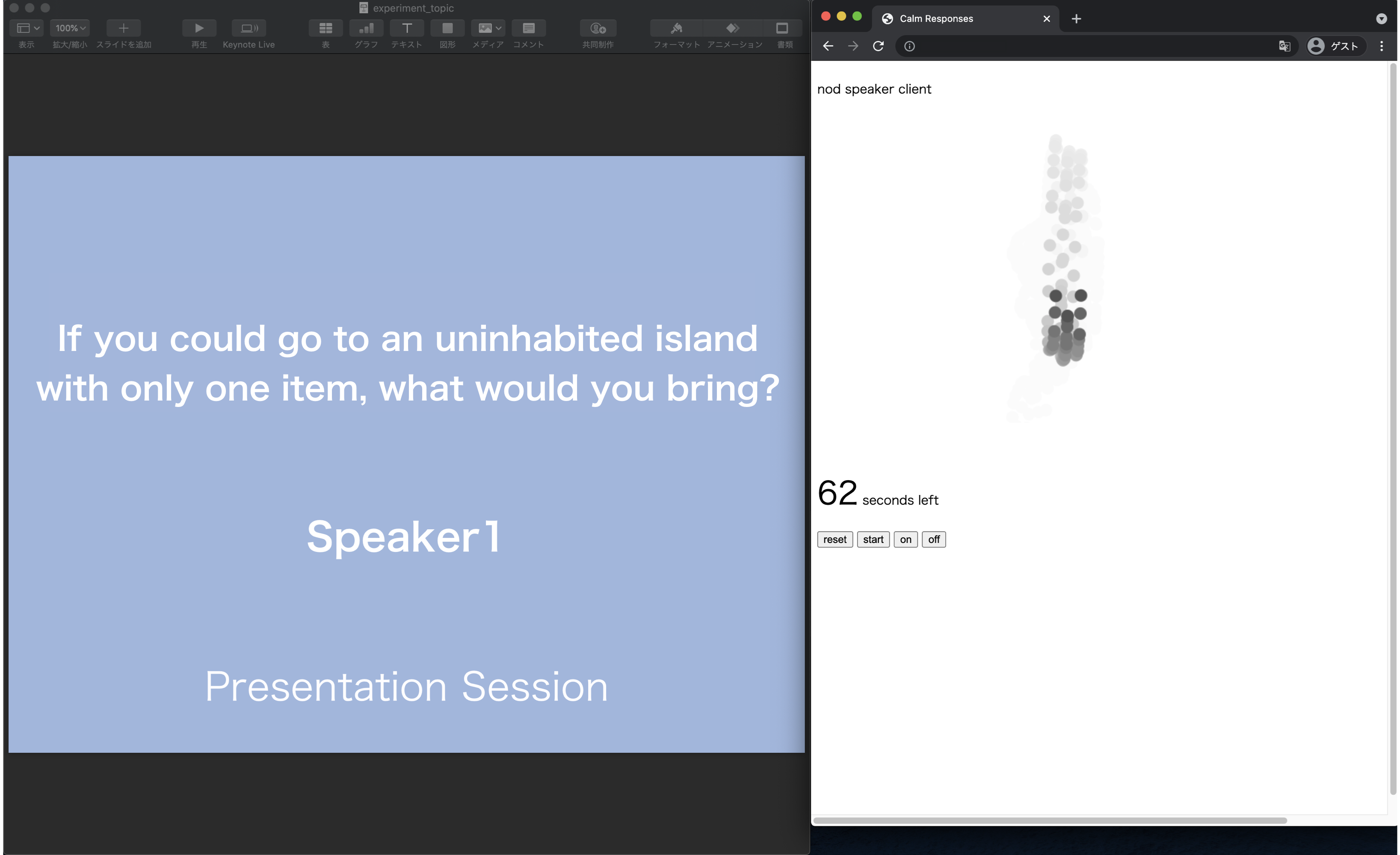}
\caption{The example shared display on Zoom during presentation in {\it condition CR-N}. While only the slide about a topic and the remaining time were presented in {\it condition B}, the head movements of audiences were also presented in {\it condition CR-N}. All participants including audiences could see them.}
\label{fig:exp2_screen}
\end{figure}

Based on the feedback received in Experiment 1, we shared head movement visualizations with both speakers and audiences on Zoom in Experiment 2. After the audiences accessed CalmResponses, they switched to Zoom, and a speaker and audiences viewed a shared screen during several presentations. Figure \ref{fig:exp2_screen} shows the shared screen on Zoom. All participants (speakers + audiences) could see the head movements' visualization, the remaining time, and a topic of the presentation in {\it condition CR-N}. In {\it condition B}, on the other hand, they could not see the visualization of the reactions. The presentation topics in both {\it condition CR-N} and {\it condition B} were: ``If you could go to an uninhabited island with only one item, what would you bring?'', ``If you could use 100 thousand dollars in only one day, what would you use it for?'', and ``If you had a time machine, what period do you want to visit?''. The speakers talked about different topics in the two presentations.

\subsection{Result}

\begin{figure*}[t]
\centering
\includegraphics[width=15cm]{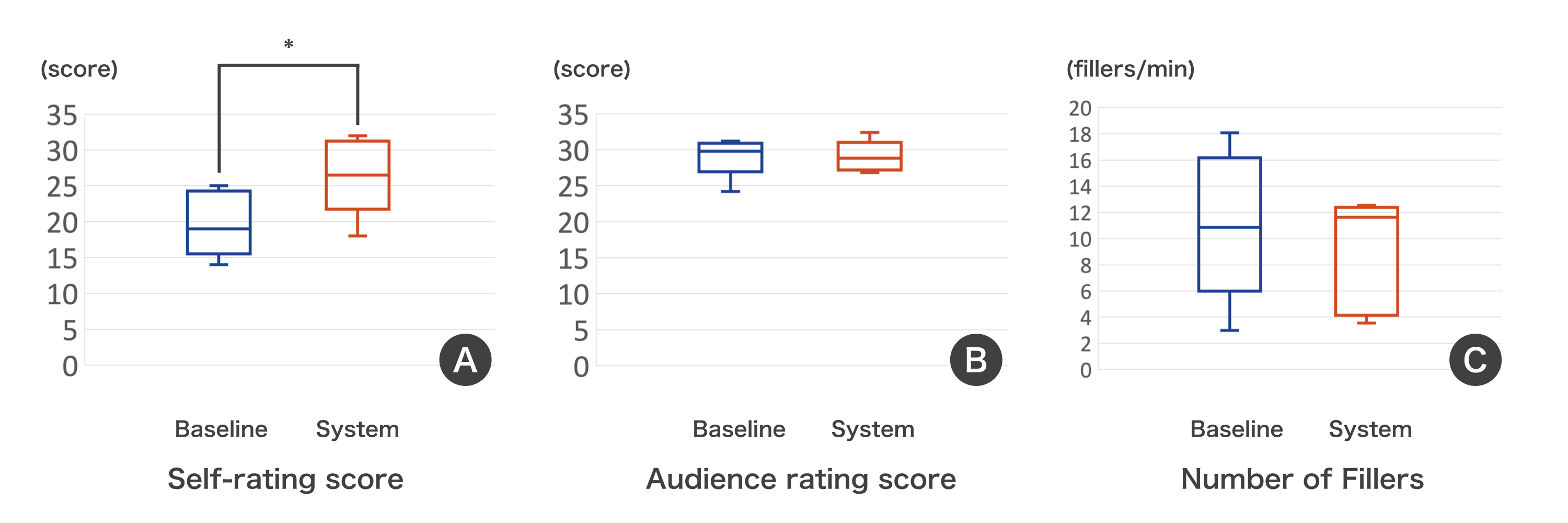}
\caption{Self-rating score (A), Audience rating score (B), Number of fillers per minute (C). p < .05 is marked as *.}
\label{fig:experiment2_rating}
\end{figure*}

\subsection*{How Did the Speakers Change Subjectively?}
\label{experiment2_result_selfrating}

The result of the paired t-test showed no significant difference between the baseline and system conditions on the State Anxiety (p = .41). Figure \ref{fig:experiment2_rating}A shows the self-evaluation scores in both {\it condition B} and {\it condition CR-N}. This score showed a significant difference between the two conditions (p = .024). This suggests that the use of the system increased speakers' self-evaluation ratings. Four out of six speakers exhibited higher self-evaluation scores in {\it condition CR-N}. In the collected comments, we found that they felt relieved when they were using CalmResponses, which led to high self-evaluation scores: {\it ``I had a calm feeling when I used the system because I was sure that the audience was listening''} [SN4]. {\it ``I was mentally relieved when I saw reactions because they expressed the agreement''} [SN6]. This feedback was consistent with the finding of Experiment 1. Another speaker implied that the higher rating was related to the expansion of the content she talked about: {\it ``I could expand on the content smoothly and speak more without thinking too much due to the feedback from the system''} [SN5]. This is consistent with the finding from prior research that audience feedback helped speakers speak more \cite{murali2021affectivespotlight}. Another speaker associated the self-rating with the fact that they could look at themselves objectively: {\it "I was able to consider the audience reactions with the system, so my self-rating score got higher"} [SN1].


However, two speakers mentioned that the system did not affect their self-evaluations much. One of them pointed out that nodding was not enough reaction to estimate audience engagement: {\it ``I could not tell whether the audience was feeling enjoyment or bored only from the nodding feedback''} [SN3]. The other participant did not observe many reactions: {\it ``Even though the head movements were visualized, there were not many nodding reactions''} [SN2]. 
Despite this, SN2 was aware of his own presentation: {\it ``I tried to improve my presentation so that audiences would react more''} [SN2]. This aspect could be considered one of the system's potential benefits.

\subsection*{How Did the Speakers Change Objectively?}
\label{experiment2_result_performance}

Figure \ref{fig:experiment2_rating}B shows the audience rating scores. There was no significant difference between the two conditions (p = .42). This result indicates that the system had a limited effect on the objective speech quality. This was also consistent with the small difference in the number of fillers per minute, as described in Figure \ref{fig:experiment2_rating}C (p = .21).

\begin{figure*}[h]
\centering
\includegraphics[width=0.6\linewidth]{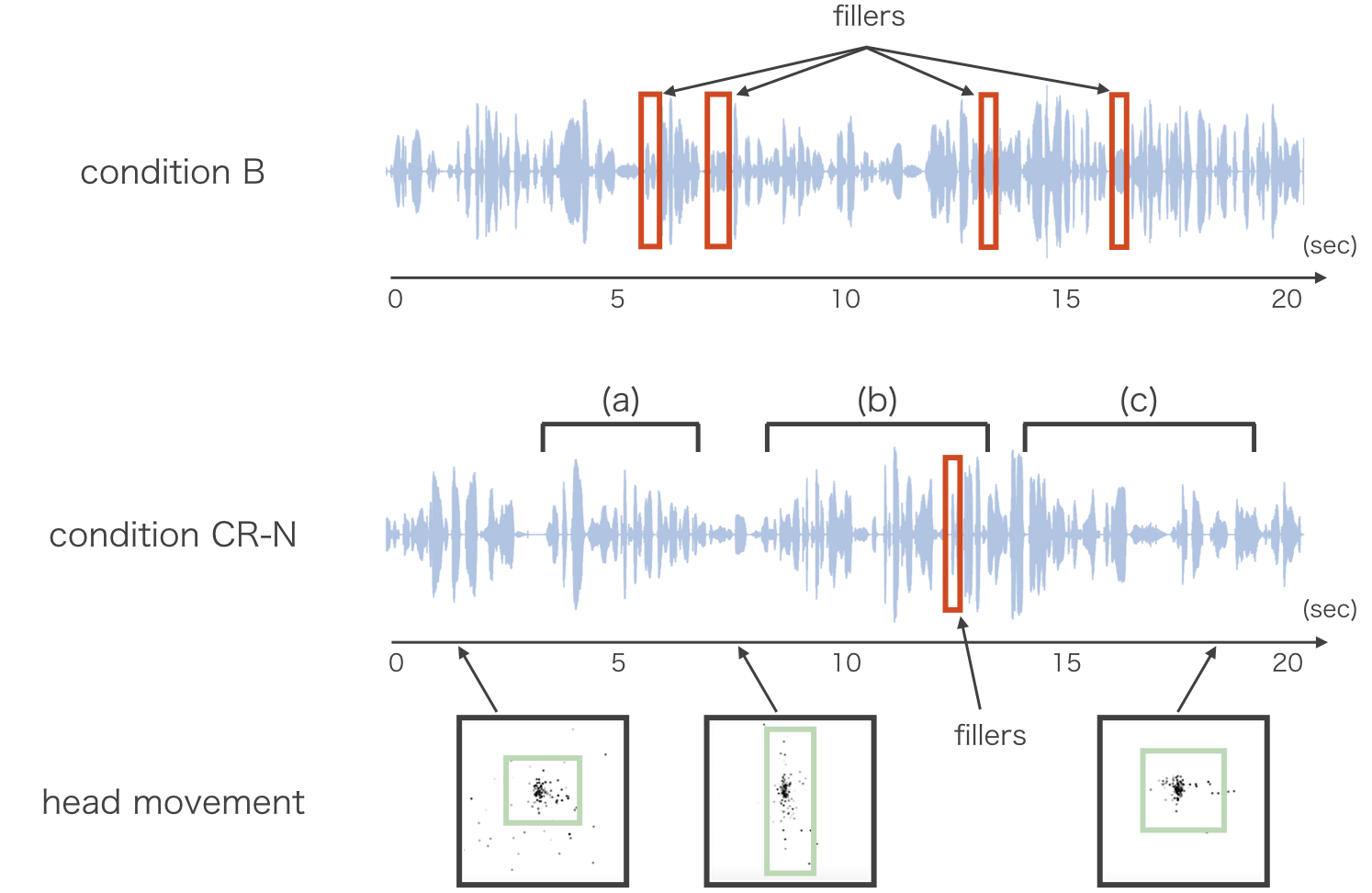}
\caption{Analysis of a speaker's 20 second speech in both conditions. We annotated filler positions manually and they are marked as orange boxes. Moreover in {\it condition CR-N}, we show how nod reactions were presented during the speech.}
\label{fig:speech_analysis}
\end{figure*}

While the number of fillers did not change much, one speaker's fillers were remarkably reduced. Figure \ref{fig:speech_analysis} shows the timeline of the middle of the 20-second speeches given by SN5 who did the first speech under {\it condition B} and the second under {\it condition CR-N}, illustrating how the speaker's number of fillers per minute was much lower in {\it condition CR-N} (4.04 fillers / min) than in {\it condition B} (8.25 fillers / min). As can be seen in this figure, in {\it condition B}, SN5 often uttered fillers, which indicated that SN5 could not put the idea into words in those periods. Moreover, SN5 did not pause much during the speech in {\it condition B}, as there was no reaction for SN5 to observe. In comparison, in {\it condition CR-N}, the speaker rarely uttered fillers.

To further explore how the speaker perceived the system, we correlated the speech with head movements visualization. The bottom of the Figure \ref{fig:speech_analysis} shows how audience head movements were displayed in {\it condition CR-N}. In this condition, SN5 gave a presentation about ``If you could go to an uninhabited island with only one item, what would you bring?''. Before (a), SN5 explained the premise of the topic, and at this time, the head movements were relatively still. When SN5 claimed that fire and water are important to life on an uninhabited island at (a), many audience members nodded. After that, SN5 paused the speech a while, presumably because SN5 wanted to observe the audience reactions. Cursors continued to move vertically, while the speakers decided to bring a tool to obtain fire or water at (b). At this point, the speaker sometimes paused the speech. From (c), SN5 was thinking about a concrete tool to bring in the island, and head movements were relatively still again. This finding implied that the audience members nodded when they heard the speaker's opinion and in that moment, the speaker used pauses to observe reactions. Although further verification is demanded, this analysis implied the efficacy of collectively presenting nodding reactions to the speakers in terms of improving their fluency of the presentation.

\section{Discussion}

\subsection{The Effect of Collective Reactions}

In the present study, we examined how using a browser-based system to monitor audience reactions affected speakers' behaviors. We found a trend in which speakers' self-rating scores were higher in {\it condition CR-N}, and the number of fillers was smaller in {\it condition CR-E}. In this section, we discuss the implications of eye gaze and nodding as audience reactions and collective visualizations based on the experimental results.

\subsubsection{{\bf Advantage of Each Reaction}}

It would be essential to take advantage of each reaction rather than comparison. In their interviews, participants suggested the suitable situations for eye gaze sharing as also described in the result of the first experiment: {\it ``I thought slide-based presentations were good for the system because visual stimuli frequently changed''} [SE5]. {\it ``The system would be more effective when speakers were using several slides''} [AE16]. In comparison, nodding might be helpful in bidirectional communication due to turn-taking and coordination functions. With regard to other reactions, speakers telling a funny story to audiences at a comedy show might want to observe smiles from facial expressions rather than eye gaze or head movements. Since we observed different results in terms of self-rating scores and the number of fillers between eye gaze and nodding, it is implied that the result will change depending on modalities and situations. It is desirable to further sophisticate the feedback for various situations given these different characteristics of different reactions.

\subsubsection{{\bf Privacy Concern}}

Although we only sent eye gaze positions and nose tip velocities through the server, privacy concerns are an inevitable issue when using webcams \cite{portnoff2015somebody}. Interviews with audience members reflected this: {\it ``I was worried a little about being watched my video''} [AE10]. {\it ``I felt I was being monitored''} [AE13]. {\it ``I was a little embarrassed when using the system''} [AN1]. These statements imply that the system invaded some audiences' privacy. However, other audiences expressed the opposite: {\it ``I was slightly embarrassed, but that was almost no problem because I thought that the system could keep anonymity''} [AE3]. {\it ``I thought that my impression would change depending on whether I could identify individuals from head movements or not. In this experiment, I was relieved that others couldn't identify individuals from the visualization''} [AN4]. These comments suggest that collectively presenting information retained anonymity to some extent because individuals could not identify who were nodding based on the cursor visualization and where others were looking based on heatmap visualization. The comments in the second experiment had a stronger tendency than the first one, which suggests that the audiences in the second experiment reported fewer privacy concerns than those in the first experiment. Anonymity is one of the important advantages of remote communication, as mentioned in \cite{hollan1992beyond}. Since privacy concerns are the main issue in the psychological and physical aspects of online sensing \cite{cooney2018pitfalls}, collective visualization will pave the way for privacy-preserving audience sensing and feedback technologies. However, in the present experiment, there was still a variation in audience perception regarding the point of privacy. This relationship between collective visualization and privacy concerns must be explored further.

\subsubsection{{\bf Co-Presence}}

Regarding the shared visualization of collective head movements in the second experiment, some audiences reported another implication. {\it ``I felt that I was involved in the presentations''} [AN2], {\it ``I was not sure the speakers' speeches were improved, but I felt a sense of unity from others' head movements visualization''} [AN6]. These comments indicate that the proposed system formed co-presence, a sense of being together psychologically \cite{bulu2012place}. This can also be related to emotional contagion \cite{hatfield1993emotional}, through which audience members are affected by one another. This is consistent with prior studies that have proposed systems of sharing visual cues with remote users to improve co-presence \cite{kim2014improving}. Co-presence has been one of the main topics in distributed communication \cite{webb2016distributed}, and many works have proposed various methods of forming co-presence \cite{jo2016effects,kim2014improving,podkosova2018co,DBLP:journals/pacmhci/ArakawaY21}. These existing methods use additional devices (e.g., HMD) or require users' intentional actions. Our proposed approach's strengths, in contrast with these existing approaches, are that it (1) requires only a commonly available device and (2) influences multiple audiences at the same time. 
Our future research will investigate whether co-presence generated by the system can improve audience performance in terms of their concentration.

\subsection{Further Applications}

\begin{figure*}[t]
\centering
\includegraphics[width=0.6\linewidth]{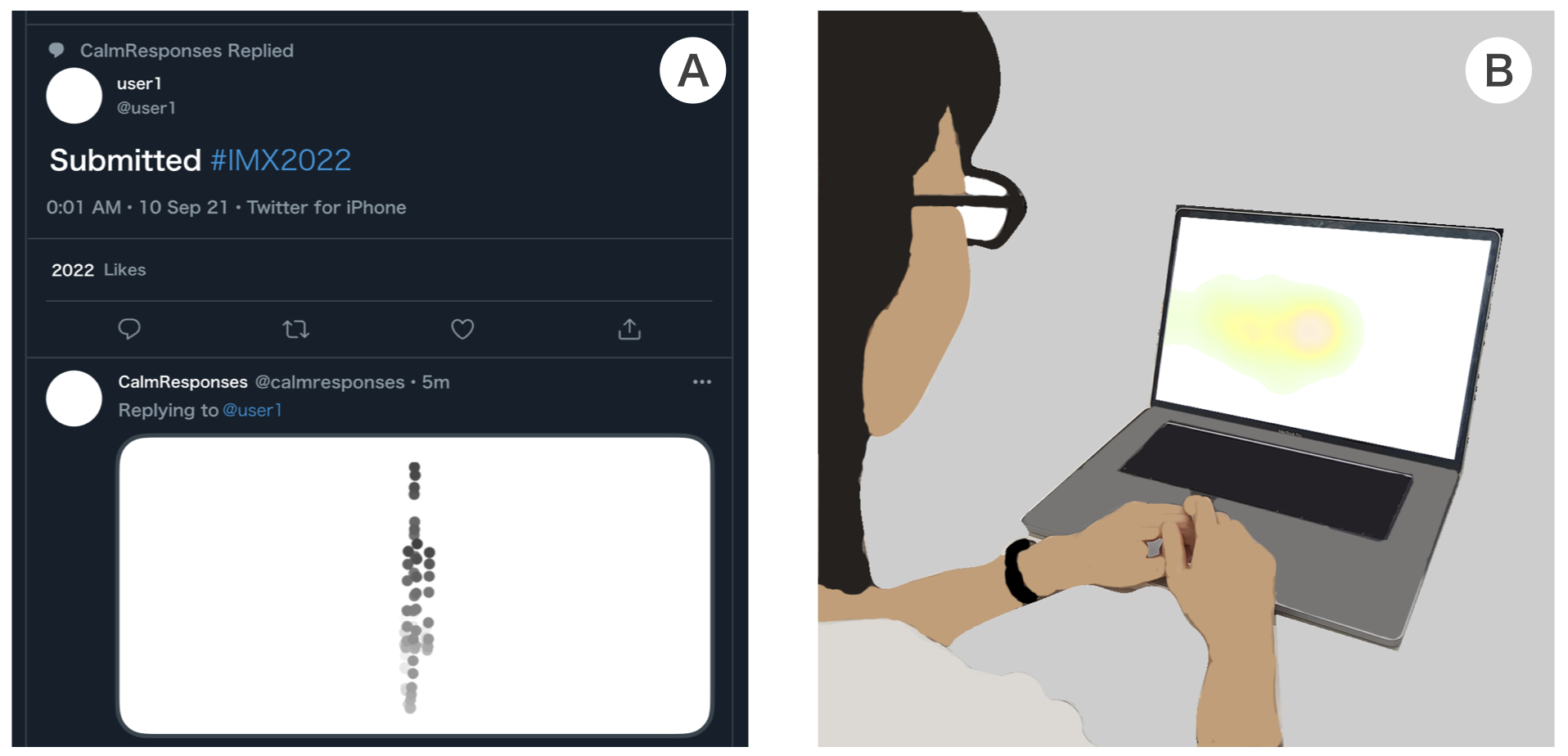}
\caption{Application scenarios in asynchronous situations. (A) Nodding Reactions on Twitter. Instead of sending two-bit information such as like or not, users can send nod reactions changing the strength of head movements to show the degree of empathy. (B) Eye gaze positions in lecture videos. Accumulated gaze positions of past learners can help new learners to observe important points of the videos quickly.}
\label{fig:application_scenario}
\end{figure*}

Regarding nod reactions, users can send reactions through the system in asynchronous situations, such as social networking services. Figure \ref{fig:application_scenario}A shows the examples of how users send nod reactions with CalmResponses and how they are visualized on Twitter. Although existing reactions on Twitter (e.g, {\it liking} or {\it retweeting a post}) are discrete, two-bit information, we can send continuous and complex reactions, such as one's degree of empathy, by changing the strength of one's head movements.

We can also leverage eye gaze reactions in asynchronous situations.
Since we can infer one's interest from eye gaze movements, collective eye gaze reactions can be used to estimate the importance of the contents in videos.
Figure \ref{fig:application_scenario}B shows an example of visualization of past learners' gaze data on a lecture video.
One of the problems with lecture videos is that learners seldom interact with their instructors and often lose their attention while watching them \cite{guo2014video,hone2016exploring}.
To overcome this problem, past learners' collective gaze visualization can help new learners to understand the important topic of the lecture videos quickly.
Since these applications do not require active participation of audiences, it will be easier to collect reactions than other systems visualizing explicit reactions and interactions of past audiences or learners \cite{kimGCLGM14,leeLCWSK15}.
Note that in these contexts, we need to consider some aspects such as note-taking and additional displays, which could invalidate the sensing method of the system.

\subsection{Limitations}

Although the present study produced some positive results, it is not without limitations. We first refer to the limitations of the system. The accuracy of the current eye-tracking module was not so high. This can be attributed to several aspects, such as the lighting conditions. Some speakers and audience members pointed out this issue: {\it ``I thought that it would be better if the accuracy was higher''} [SE3]. {\it ``There was room for improvement in accuracy''} [AE10]. We will address this limitation to use other eye-tracking modules with higher accuracy such as \cite{Zhang2020}.

In terms of the limitations of the experiments, we did not compare eye gaze and nod reactions with explicit reactions such as texts or emoticons. There is little research on the usage of these reactions in synchronous distributed communication compared to asynchronous communication. Moreover, we had limited sample sizes, especially the speaker sample size. We need to investigate further the difference between explicit reactions and eye gaze or nodding in the context of a larger-scale experiment. Another limitation is a relatively short time (approximately two minutes) and small number of topics (five topics in total) for presentation used in our experiments, compared to existing research that conducted similar experiments \cite{murali2021affectivespotlight, trinh2017robocop}. This choice has both advantages and disadvantages. On one hand, since speakers do not need much time to prepare for short speech, it could induce spontaneous and natural presentations. On the other hand, it can not be denied that approximately two minute presentations are not enough for audiences to properly assess them. Furthermore, there are various topics for actual communications, and we still do not know whether the system can be applied to other topics. In future work, we will introduce the system to in-the-wild situations such as lectures or webinars and evaluate its efficacy with longer presentations with various topics.

\section{Conclusion}

In this study, we proposed a browser-based system that displays audience eye gaze positions and head movements collectively to speakers in synchronous remote communication using a built-in webcam. We conducted two experiments to evaluate the effectiveness of the proposed system. We found that the system helped speakers reduce the number of fillers during their presentation and raised their self-rating scores. Based on our results and findings, we discussed the potential benefits of the system in terms of privacy concerns, co-presence of the audience. We also discussed the application scenarios of the system beyond the synchronous online communication. We believe that our approach and findings will enhance the experience of future remote communication by leveraging its benefits for both speakers and audiences.

\begin{acks}
This work was supported by the commissioned research by National Institute of Information and Communications Technology (NICT) Japan, JST CREST Grant Number JPMJCR17A3, and JST Moonshot R\&D Grant Number JPMJMS2012.
\end{acks}

\bibliographystyle{ACM-Reference-Format}
\bibliography{sample-manuscript}

\end{document}